\documentclass[fleqn]{2017SCGE}
\begin{document}

\ensubject{subject}

\ArticleType{Article}
\Year{2020}
\Month{January}
\Vol{63}
\No{1}
\DOI{10.1007/s11433-019-9422-1}
\ArtNo{212011}
\ReceiveDate{April 22, 2019}
\AcceptDate{April 30, 2019}
\OnlineDate{October 8, 2019}

\title{Ground state properties and potential energy surfaces of $^{270}$Hs 
       from multidimensionally-constrained relativistic mean field model}
      {Ground state properties and potential energy surfaces of $^{270}$Hs 
       from multidimensionally-constrained relativistic mean field model}

\author[1,2]{Xu MENG}{}
\author[3]{Bing-Nan LU}{}
\author[1,2,4,5]{Shan-Gui ZHOU}{sgzhou@itp.ac.cn}

\AuthorMark{Meng X, Lu B N and Zhou S G}

\AuthorCitation{Meng X, Lu B N and Zhou S G}

\address[1]{CAS Key Laboratory of Theoretical Physics, 
            Institute of Theoretical Physics, Chinese Academy of Sciences, 
            Beijing 100190, China}
\address[2]{School of Physical Sciences, 
            University of Chinese Academy of Sciences, Beijing 100049, China}
\address[3]{Facility for Rare Isotope Beams and Department of Physics and Astronomy, 
            Michigan State University, MI 48824, USA}
\address[4]{Center of Theoretical Nuclear Physics, 
            National Laboratory of Heavy Ion Accelerator, Lanzhou 730000, China}
\address[5]{Synergetic Innovation Center for Quantum Effects and Application, 
            Hunan Normal University, Changsha 410081, China}


\abstract{%
We study the ground state properties, potential energy curves and potential energy
surfaces of the superheavy nucleus $^{270}$Hs by using the multidimensionally-constrained 
relativistic mean-field model with the effective interaction PC-PK1. 
The binding energy, size and shape as well as single particle shell
structure corresponding to the ground state of this nucleus are obtained. 
$^{270}$Hs is well deformed and exhibits deformed doubly magic feature in 
the single neutron and proton level schemes. 
One-dimensional potential energy curves and two-dimensional potential energy 
surfaces are calculated for $^{270}$Hs with various spatial symmetries imposed.
We investigate in detail the effects of the reflection asymmetric and triaxial 
distortions on the fission barrier and fission path of $^{270}$Hs.
When the axial symmetry is imposed, the reflection symmetric and reflection 
asymmetric fission barriers both show a double-hump structure and the former is higher. 
However, when triaxial shapes are allowed the reflection symmetric barrier is 
lowered very much and then the reflection symmetric fission path becomes favorable.
}



\maketitle



\begin{multicols}{2}

\section{Introduction}\label{section1}

Exploration of the charge and mass limits of atomic nuclei is at the forefront 
of modern nuclear physics research \cite{Hamilton2013_ARNPS63-383,
Nazarewicz2018_NatPhys14-537,Giuliani2019_RMP91-011001}.
When described as charged liquid drops, nuclei with $Z \ge 104$ would not
exist because the fission barriers would disappear. 
Quantum shell effects may cause additional binding in some of these nuclei.
Consequently, the existence of an ``island of stability'' of superheavy nuclei 
(SHN) was predicted in 1960s because of this extra binding
\cite{Myers1966_NP81-1,Wong1966_PL21-688,
Sobiczewski1966_PL22-500,Meldner1967_ArkivF36-593,Mosel1969_ZP222-261,
Nilsson1969_NPA131-1}.
Since then, study on SHN has progressed very much.
Until now, superheavy elements (SHE) with 
$Z \le 118$ have been synthesized
\cite{Hofmann2000_RMP72-733,Morita2015_NPA944-30,Oganessian2017_PS92-023003}.
\Authorfootnote

\noindent 
However, the island is still not located and various predictions of its center 
have been made \cite{Myers1966_NP81-1,Wong1966_PL21-688,
Sobiczewski1966_PL22-500,Meldner1967_ArkivF36-593,Mosel1969_ZP222-261,
Nilsson1969_NPA131-1,Rutz1997_PRC56-238,Zhang2005_NPA753-106,
Sobiczewski2007_PPNP58-292,Zhou2011_NSC2010_279,Li2014_PLB732-169,
Mo2014_PRC90-024320,Afanasjev2018_PLB782-533,Agbemava2019_PRC99-034316}.
Contrary to the fact that the island of stability is very elusive and the 
predicted centers of this island are far beyond the current experimental ability,
the existence of a ``shallow'' of SHN, which is closer to the continent of 
stable nuclei, has been theoretically and experimentally well established.
This shallow was predicted to comprise deformed SHN and to be centered around
$Z=108$ and $N=162$ \cite{Moeller1974_NPA229-292,Cwiok1983_NPA410-254,
Patyk1989_NPA502-591,Patyk1991_NPA533-132,Smolanczuk1995_PRC52-1871}.
Many SHN on the shallow have been synthesized, including $^{270}_{108}$Hs$_{162}$ \cite{Dvorak2006_PRL97-242501,Oganessian2013_PRC87-034605}.
The deformed doubly magic SHN $^{270}$Hs sitting at the center of the shallow 
has become one of the most important examples for the study on the structure and 
fission properties of SHN.

Study on the ground state and fission barrier is essential in learning 
the stability properties, decay and fission dynamics of SHN.
Although the most relevant shape degree of freedom involved in nuclear fission 
is the axial quadrupole deformation $\beta_{20}$ corresponding to the elongation 
of the whole nucleus along its symmetry axis, other shape degrees of freedom 
may also play crucial roles during the fission process 
\cite{Pashkevich1969_NPA133-400,Moeller1973_IAEA-SM-174-202,
Rutz1995_NPA590-680,
Robledo2008_IJMPE17-204,
Kowal2010_PRC82-014303,
Li2010_PRC81-064321,
Abusara2010_PRC82-044303,
Staszczak2011_IJMPE20-552,
Royer2012_PRC86-044326,
Lu2012_PRC85-011301R,
Warda2012_PRC86-014322,Lu2014_PRC89-014323,Zhou2016_PS91-063008}. 
Therefore examining the multidimensional potential energy surfaces (PESs) for 
a nucleus in question is important for obtaining detailed information 
about the fission properties.

We have developed multidimensionally-constrained (MDC) covariant density 
functional theories (CDFTs) to calculate and study PESs for SHN and heavy nuclei 
\cite{Lu2012_PRC85-011301R,Lu2014_PRC89-014323,
Zhou2016_PS91-063008,Zhao2017_PRC95-014320}.
In these theories, the shape degrees of freedom $\beta_{\lambda\mu}$ with $\mu$ 
being even numbers are self-consistently included, such as $\beta_{20}$, $\beta_{22}$, 
$\beta_{30}$, $\beta_{32}$, $\beta_{40}$, $\beta_{42}$ and $\beta_{44}$. 
Either the BCS approach or the Bogoliubov transformation has been implemented 
for considering pairing correlations.
The MDC-CDFT with the BCS approach used for treating pairing correlations is 
known as the MDC relativistic mean field (RMF) model, whereas that with the 
Bogoliubov transformation is known as the MDC relativistic Hartree--Bogoliubov 
(RHB) model.
The MDC-RMF model has been used to investigate fission barriers and PESs 
of actinide nuclei; in particular, a three-dimensional PES has been obtained 
for $^{240}$Pu \cite{Lu2012_PRC85-011301R,Lu2014_PRC89-014323}. 
It was found that besides the reflection asymmetric octupole shape, the triaxial
deformation plays a crucial role on the second fission barriers in actinide 
nuclei.

In this work, we use the MDC-RMF model to study the ground state properties and 
PESs of the doubly magic deformed nucleus $^{270}$Hs. 
This is the first thorough study on SHN with MDC-CDFTs.
We present the first PES for $^{270}$Hs considering both triaxiality
and reflection asymmetry.
The remainder of this paper is organized as follows. 
The MDC-RMF model will be introduced in \autoref{sec:model}. 
The results and discussions are presented in \autoref{sec:results}. 
Finally in \autoref{sec:summary}, we will summarize this study and 
provide some perspectives on the study on SHN with MDC-CDFTs.

\section{The multidimensionally-constrained relativistic mean field model
\label{sec:model}}

CDFT is one of the most successful self-consistent approaches which
has been extensively used to study the properties
of atomic nuclei throughout the nuclear chart
\cite{Serot1986_ANP16-1,Reinhard1989_RPP52-439,%
Ring1996_PPNP37-193,Bender2003_RMP75-121,Vretenar2005_PR409-101,%
Meng2006_PPNP57-470,Paar2007_RPP70-691,Niksic2011_PPNP66-519,%
Liang2015_PR570-1,Meng2015_JPG42-093101,Meng2016_RDFNS}.
The formalism and applications of the CDFT can be found in Ref.~\cite{Meng2016_RDFNS}.
The details of the MDC-CDFTs have been given in Refs.~\cite{Lu2014_PRC89-014323,
Zhou2016_PS91-063008,Zhao2017_PRC95-014320}.
For completeness, here we present briefly the formalism of MDC-CDFTs.

In the CDFT, a nucleus is described to be a composite of $A$ nucleons which 
interact through the exchanges of mesons and photons or contact interactions (point-couplings).
In this paper we focus on the CDFT with nonlinear point couplings (NL-PC).
The time-reversal symmetry is assumed for nuclei in question.

The NL-PC Lagrangian reads,
\begin{eqnarray}
 \mathcal{L} = \bar{\psi}(i\gamma_{\mu}\partial^{\mu}-M)\psi
              -\mathcal{L}_{{\rm lin}}
              -\mathcal{L}_{{\rm nl}}
              -\mathcal{L}_{{\rm der}}
              -\mathcal{L}_{{\rm Cou}},
\end{eqnarray}
where the linear, nonlinear and derivative coupling terms
and the Coulomb term are given respectively as follows,
\begin{eqnarray}
 \mathcal{L}_{{\rm lin}} & = & \frac{1}{2} \alpha_{S} \rho_{S}^{2}
                              +\frac{1}{2} \alpha_{V} \rho_{V}^{2}
                              +\frac{1}{2} \alpha_{TS} \vec{\rho}_{TS}^{2}
                              +\frac{1}{2} \alpha_{TV} \vec{\rho}_{TV}^{2} ,
\label{eq:L_lin}
 \\
 \mathcal{L}_{{\rm nl}}  & = & \frac{1}{3} \beta_{S} \rho_{S}^{3}
                              +\frac{1}{4} \gamma_{S}\rho_{S}^{4}
                              +\frac{1}{4} \gamma_{V}[\rho_{V}^{2}]^{2} ,
\label{eq:L_nl}
 \\
 \mathcal{L}_{{\rm der}} & = & \frac{1}{2} \delta_{S}[\partial_{\nu}\rho_{S}]^{2}
                              +\frac{1}{2} \delta_{V}[\partial_{\nu}\rho_{V}]^{2}
                              +\frac{1}{2} \delta_{TS}[\partial_{\nu}\vec{\rho}_{TS}]^{2}
\label{eq:L_der}
 \\
 &  & \mbox{}                 +\frac{1}{2} \delta_{TV}[\partial_{\nu}\vec{\rho}_{TV}]^{2} ,
 \nonumber \\
 \mathcal{L}_{{\rm Cou}} & = & \frac{1}{4} F^{\mu\nu} F_{\mu\nu}
                             +e\frac{1-\tau_{3}}{2} A_{0} \rho_{V} .
\label{eq:L_Cou}
\end{eqnarray}
$M$ is the nucleon mass and $\alpha_{S}$, $\alpha_{V}$, $\alpha_{TS}$,
$\alpha_{TV}$, $\beta_{S}$, $\gamma_{S}$, $\gamma_{V}$, $\delta_{S}$,
$\delta_{V}$, $\delta_{TS}$ and $\delta_{TV}$ are the coupling constants 
in different channels. 
$\rho_{S}$ and $\vec{\rho}_{TS}$ are the isoscalar and isovector densities;
$\rho_{V}$ and $\vec{\rho}_{TV}$ are the time-like components of isoscalar
and isovector currents.

Applying the mean field and no sea approximations, the Dirac equation for 
nucleons is derived with the Slater determinant used as the trial wave function,
\begin{eqnarray}
 \hat{h}\psi_{k}(\bm{r}) = \epsilon_{k} \psi_{k}(\bm{r}) ,
 \label{eq:Diracequation}
\end{eqnarray}
where
\begin{eqnarray}
 \hat{h} = \bm{\alpha} \cdot \bm{p}
         + \beta \left[ M+S(\bm{r}) \right]
         + V(\bm{r}) .
 \label{eq:dirac}
\end{eqnarray}
The scalar potential $S(r)$ and vector potential $V(r)$ are determined by the 
various scalar and vector densities given in Eqs.~(\ref{eq:L_lin}--\ref{eq:L_Cou}).

One can solve the Dirac equation in several different harmonic oscillator bases 
\cite{Gambhir1990_APNY198-132,Ring1997_CPC105-77,Afanasjev2000_NPA676-196,
Geng2007_CPL24-1865,Zhang2010_PRC81-034302,
Wang2018_SciChinaPMA61-082012,Qi2019_SciChinaPMA62-012012,Xia2019_SciChinaPMA62-042011}.
In the MDC-CDFTs, an axially deformed harmonic oscillator (ADHO) basis
\cite{Gambhir1990_APNY198-132,Ring1997_CPC105-77}
is used to expand the single particle Dirac wave functions.
The ADHO basis consists of the eigenstates of the Schr\"odinger equation,
\begin{eqnarray}
 \left[-\frac{\hbar^{2}}{2M}\nabla^{2}+V_{B}(z,\rho)\right]\Phi_{\alpha}(\bm{r}\sigma)
 & = & E_{\alpha}\Phi_{\alpha}(\bm{r}\sigma) ,
 \label{eq:BasSchrodinger-1}
\end{eqnarray}
where $\bm{r} = (z,\rho)$ with $\rho=\sqrt{x^2+y^2}$ and
\begin{eqnarray}
 V_{B}(z,\rho) = \frac{1}{2} M ( \omega_{\rho}^{2} \rho^{2} + \omega_{z}^{2} z^{2})
 ,
\end{eqnarray}
is the ADHO potential with $\omega_{\rho}$ ($\omega_{z}$) being
the oscillator frequency perpendicular to (along) the symmetry axis.
The solution of Eq.~(\ref{eq:BasSchrodinger-1}) is obtained as
\begin{eqnarray}
 \Phi_{\alpha}(\bm{r}\sigma)
 =
 C_{\alpha} \phi_{n_{z}}(z) R_{n_{\rho}}^{m_{l}}(\rho)
 \frac{1}{\sqrt{2\pi}} e^{im_{l}\varphi}
 \chi_{m_s}(\sigma),
 \label{eq:HO}
\end{eqnarray}
where a complex number $C_{\alpha}$ is introduced for convenience.
$\chi_{m_{s}}$ is a two-component spinor.
$\phi_{n_{z}}(z)$ and $R_{n_{\rho}}^{m_{l}}(\rho)$ are the harmonic oscillator wave functions,
\begin{eqnarray}
 \phi_{n_{z}}(z) & = &
 \frac{1}{\sqrt{b_{z}}} \frac{1}{\pi^{{1}/{4}} \sqrt{2^{n_{z}}n_{z}!}}
  H_{n_{z}}\left(\frac{z}{b_{z}}\right)
  e^{-{z^{2}}/{2b_{z}}} ,
  \\
  R_{n_{\rho}}^{m_{l}}(\rho) & = &
  \frac{1}{b_{\rho}} \sqrt{\frac{2n_{\rho}!}{(n_{\rho}+|m_{l}|)!}}
  \left( \frac{\rho}{b_{\rho}} \right)^{|m_{l}|}
  \nonumber \\
  & & \mbox{\hspace{2cm}} \times
  L_{n_{\rho}}^{|m_{l}|}
  \left( \frac{\rho^{2}}{b_{\rho}^{2}} \right)
  e^{ -{\rho^{2}}/{2b_{\rho}^{2}} }
  .
\end{eqnarray}
The oscillator lengths $b_{z}$ and $b_{\rho}$ are related to the frequencies as
$b_{z}=1/\sqrt{M\omega_{z}}$ and $b_{\rho}=1/\sqrt{M\omega_{\rho}}$.

These basis states
are eigenstates of 
$\hat{j}_{z}$ with eigenvalues $K_\alpha=m_{l}+m_{s}$.
The deformation of the basis $\beta_{{\rm basis}}$ is defined through the relations
$\omega_{z}=\omega_{0}\exp\left(-\sqrt{{5}/{4\pi}}\beta_{{\rm basis}}\right)$
and $\omega_{\rho}=\omega_{0}\exp\left(\sqrt{{5}/{16\pi}}\beta_{{\rm basis}}\right)$,
where $\omega_{0}=(\omega_{z}\omega_{\rho}^{2})^{{1}/{3}}$ is the frequency of
the corresponding spherical oscillator potential.

When solving the RMF equations, the Dirac spinor 
is expanded in terms of the complete basis $\{ \Phi_{\alpha}(\bm{r}\sigma) \}$ as,
\begin{eqnarray}
 \psi_{i}(\bm{r}\sigma) =
 \left(
  \begin{array}{c}
   \sum_{\alpha}f_{i}^{\alpha} \Phi_{\alpha}(\bm{r}\sigma) \\
   \sum_{\alpha}g_{i}^{\alpha} \Phi_{\alpha}(\bm{r}\sigma)
  \end{array}
 \right),
\label{eq:spwaveexpansion}
\end{eqnarray}
where 
$\alpha=\{n_{z},n_{\rho},m_{l},m_{s}\}$ and $f_{i}^{\alpha}$ and
$g_{i}^{\alpha}$ are the expansion coefficients to be determined.
Following Ref.~\cite{Warda2002_PRC66-014310},
the ADHO basis is truncated with 
$[ n_{z}/f_{z}+(2n_{\rho}+|m_l|)/f_{\rho} ] \le N_f$
for the upper (large) component of the Dirac spinor 
where
$f_{z}=\max(b_{z}/b_{0},1)$ and $f_{\rho}=\max(b_{\rho}/b_{0},1)$ are constants
and 
$b_{0} = 1/\sqrt{M\omega_0}$ is the spherical harmonic oscillator length.
For the expansion of the lower (small) component of the Dirac spinor,
the truncation of $N_g = N_f+1$ is made to avoid the numerical instability.

In the MDC-CDFTs, 
we assume that the nuclear potentials and densities are
invariant under the following operations:
the reflection with respect to the $y$-$z$ plane ($\hat{S}_x$),
the reflection with respect to the $x$-$z$ plane ($\hat{S}_y$) and
the rotation of 180$^\circ$ with respect to the $z$ axis ($\hat{S}$), i.e.,
\begin{eqnarray}
 \hat{S}_x \phi(x,y,z)  &=&  \phi(-x, y,z) , \label{eq:V4_1} \\
 \hat{S}_y \phi(x,y,z)  &=&  \phi( x,-y,z) , \label{eq:V4_2} \\
 \hat{S}   \phi(x,y,z)  &=&  \phi(-x,-y,z) . \label{eq:V4_3} 
\end{eqnarray}
These three operations and the identity $\hat{I}$ form the $V_4$ group.

The pairing correlations are crucial in open shell nuclei.
It has been shown that fission barriers are influenced very much by the 
pairing correlations \cite{Karatzikos2010_PLB689-72}.
In the MDC-RMF model \cite{Lu2014_PRC89-014323}, we employed the BCS approach with a separable pairing force of finite-range \cite{Tian2006_CPL23-3226,Tian2009_PLB676-44,Tian2009_PRC79-064301}.

To obtain a PES, i.e., the energy of a nucleus as a function of various shape 
degrees of freedom $E = E( \{ \beta_{\lambda\mu} \})$,
one can perform constraint calculations \cite{Ring1980}.
A modified linear constraint method has been proposed 
\cite{Lu2012_PRC85-011301R,Lu2014_PRC89-014323}
in which the Routhian is calculated as,
\begin{eqnarray}
 E^{\prime} = E_{{\rm RMF}} +
              \sum_{\lambda\mu} \frac{1}{2} C_{\lambda\mu}Q_{\lambda\mu} .
\end{eqnarray}
In the $(n+1)$th iteration, the variable $C_{\lambda\mu}$ is determined by,
\begin{eqnarray}
 C_{\lambda\mu}^{(n+1)} =
 C_{\lambda\mu}^{(n)} +
  k_{\lambda\mu} \left( \beta_{\lambda\mu}^{(n)} - \beta_{\lambda\mu} \right),
\end{eqnarray}
where $\beta_{\lambda\mu}$ is the desired deformation,
$k_{\lambda\mu}$ is a constant and $C_{\lambda\mu}^{(n)}$ is the value in the $n$th iteration.

The RMF equations are solved iteratively. After a desired accuracy is
achieved, we can calculate various physical quantities.
For example, the intrinsic multipole moments are calculated from the density by
\begin{eqnarray}
 Q_{\lambda\mu} = \int d^{3}\bm{r} \rho_{V}(\bm{r}) r^{\lambda} Y_{\lambda\mu}(\Omega),
\end{eqnarray}
where $Y_{\lambda\mu}(\Omega)$ is the spherical harmonics.
The deformation parameter $\beta_{\lambda\mu}$ is obtained from
the corresponding multipole moment by
\begin{eqnarray}
 \beta_{\lambda\mu} = \frac{4\pi} {3nR^{\lambda}} Q_{\lambda\mu},
\end{eqnarray}
where $R= r_0 A^{{1}/{3}}$ is the radius of the nucleus, the
parameter $r_0=1.2$~fm
and $n$ represents proton, neutron or nucleon numbers $Z$, $N$ or $A$.
Applying the operations Eqs.~(\ref{eq:V4_1}--\ref{eq:V4_3}) to the densities, 
one gets under the assumption of the $V_4$ symmetry, 
\begin{eqnarray}
 \beta_{\lambda\mu} =        \beta_{\lambda\bar\mu}
 = (-1)^\mu \beta_{\lambda\bar\mu}.
\end{eqnarray}
Therefore $\beta_{\lambda\mu} = 0$ when $\mu$ is an odd number.
Consequently, the four deformations of the lowest order allowed in the 
MDC-RMF model are $\beta_{20}$, $\beta_{22}$, $\beta_{30}$ and $\beta_{32}$.

The MDC-CDFTs have been applied to the study on fission barriers 
and PESs of actinide nuclei \cite{Lu2012_PRC85-011301R,Lu2014_PRC89-014323}, 
the third minima in PESs of light actinides \cite{Zhao2015_PRC91-014321}, 
the non-axial octupole $Y_{32}$ correlations in $N = 150$ isotones 
\cite{Zhao2012_PRC86-057304} and Zr isotopes \cite{Zhao2017_PRC95-014320}, 
axial octupole correlations in M$\chi$D \cite{Liu2016_PRL116-112501} 
and Ba isotopes \cite{Chen2016_PRC94-021301R} and 
shapes of hypernuclei \cite{Lu2011_PRC84-014328,Lu2014_PRC89-044307}. 
Based on the PESs from MDC-CDFTs, the dynamics of spontaneous 
and induced fissions in actinide nuclei has been studied 
\cite{Zhao2015_PRC92-064315,Zhao2016_PRC93-044315,Zhao2019_PRC99-014618,
Zhao2019_arXiv1902.09535}.
In this work, the MDC-RMF model is used to make the first thorough study on 
SHN, namely, the ground state properties and PESs 
of the doubly magic deformed nucleus $^{270}$Hs. 

\section{Results and discussions} \vspace*{-1mm}
\label{sec:results}

\subsection{Numerical details
\label{sec:numerical}}

As mentioned in \autoref{sec:model}, a truncation 
$[ n_{z}/f_{z}+(2n_{\rho}+|m_l|)/f_{\rho} ] \le N_f$
is made on the ADHO basis.
When studying the ground state properties and one-dimensional potential energy 
curves (PECs), we choose $N_f=20$. 
Such a truncation is also made when the two-dimensional PESs are calculated 
with the axial symmetry imposed.
For the two-dimensional PESs with both triaxial and reflection
asymmetric deformations considered, 
a smaller basis with $N_f=16$ is adopted
because of the limitation of computational capabilities. 
We have checked that in the deformation range we are interested in, $N_f=20$
(16) can give an accuracy of 0.1 (0.5) MeV in the total binding energy.
Generally speaking, such accuracies warrant the conclusions drawn in this
work concerning the effects of the reflection asymmetric and triaxial 
deformations on the fission barriers and fission paths of $^{270}$Hs.

In the particle-hole (ph) channel, we use the effective interaction PC-PK1 
which was determined by fitting to observables of 60 selected spherical nuclei 
including binding energies, charge radii and empirical pairing gaps 
\cite{Zhao2010_PRC82-054319}. 
As reviewed in Ref.~\cite{Zhao2019_IJMPE27-1830007}, this empirical functional
has been very successful in describing nuclear ground state properties,
e.g., the Coulomb displacement energies between mirror nuclei 
\cite{Sun2011_SciChinaPMA54-210},
nuclear binding energies \cite{Zhao2012_PRC86-064324,
Lu2015_PRC91-027304,Xia2018_ADNDT121-122-1}
and quadrupole moments \cite{Zhao2014_PRC89-011301R,Yordanov2016_PRL116-032501,
Haas2017_EPL117-62001} and
phase transitions in the nuclear shape \cite{Quan2018_PRC97-031301R},
and low-energy excited states including
nuclear chiral rotations \cite{Zhao2017_PLB773-1}
and magnetic and antimagnetic rotations \cite{Zhao2011_PLB699-181,
Zhao2011_PRL107-122501,Meng2013_FrontPhys8-55,Peng2015_PRC91-044329,
Meng2016_PS91-053008}.
PC-PK1 has also been used to study properties of SHN \cite{Zhang2013_PRC88-054324,
Agbemava2015_PRC92-054310,Li2015_FrontPhys10-268}.

In the separable pairing force of finite range used in the particle-particle (pp)
channel, there are two parameters,
the pairing strength $G$ and the effective range of the pairing force $a$.
They have been fixed by reproducing the density dependence of the pairing gap 
of symmetric nuclear matter at the Fermi surface calculated with 
the Gogny force D1S: $G=G_0 = 728$ MeV$\cdot$fm$^3$ and $a=0.644$ fm
\cite{Tian2009_PLB676-44,Tian2009_PRC80-024313}.
In the present work, a fine-tuning of the pairing strength $G$ has been made and
$G/G_0=1.1$ is used which can reproduce the odd-even mass staggering of 
$N=162$ isotones and Hs isotopes.

\subsection{Ground state properties of $^{270}$Hs
\label{sec:ground}}\vspace*{-1mm}

As a typical doubly magic deformed nucleus in the superheavy mass region,
$^{270}$Hs has been studied extensively and many of the studies focused on
the ground state properties. 
In \autoref{table:g.s.&comp} our results for the bulk properties of the ground 
state, including the binding energy $E_\mathrm{B}$, 
deformation parameter $\beta_{2}$ and root mean square (rms) charge radius
$R_\mathrm{c}$, are compared with results from other models 
as well as the empirical value from the latest ``Atomic Mass Evaluation'' 
(AME2016) \cite{Audi2017_ChinPhysC41-030001,
Huang2017_ChinPhysC41-030002,Wang2017_ChinPhysC41-030003}.

\begin{table}[H]
\footnotesize
\begin{threeparttable}
\caption{Ground state properties, including the binding energy $E_\mathrm{B}$, 
deformation parameter $\beta_{2}$ and rms charge radius
$R_\mathrm{c}$, obtained from the MDC-RMF model.
Results from other models and AME2016 are included for comparison.
See text for more details.
\label{table:g.s.&comp}}
\doublerulesep 0.1pt \tabcolsep 8pt 
\begin{tabular}{llll}
\toprule
 Model        & $E_\mathrm{B}$ (MeV) & $\beta_{2}$ & $R_\mathrm{c}$ (fm) \\ 
\hline
 MDC-RMF (PC-PK1) 
              & 1967.40     & 0.261       & 6.167                 \\
 AME2016 \cite{Audi2017_ChinPhysC41-030001,Huang2017_ChinPhysC41-030002,
               Wang2017_ChinPhysC41-030003}      
              & 1969.65     &             &                       \\ 
 MMM \cite{Patyk1991_NPA533-132}
              & 1969.20     & 0.229       &                       \\ 
 RMF (TMA) \cite{Ren2002_PRC65-051304R,Ren2002_PRC66-064306} 
              & 1971.80     & 0.22        & 6.152                 \\ 
 RMF (NLZ2) \cite{Ren2002_PRC65-051304R,Ren2002_PRC66-064306} 
              & 1969.22     & 0.274       & 6.251                 \\ 
 RMF (TMA) \cite{Geng2005_PTP113-785,Geng2006_PhD-Thesis} 
              & 1971.93     & 0.222       & 6.142                 \\ 
 HFB-24 \cite{Goriely2013_PRC88-024308} 
              & 1968.45     & 0.26        &                       \\ 
 RMF (NL3)  \cite{Zhang2012_PRC85-014325}            
              & 1974        & 0.26        &                       \\ 
 WS4 \cite{Wang2014_PLB734-215}          
              & 1970.27     & 0.217       &                       \\
 FRDM (2012) \cite{Moeller2016_ADNDT109-110-1} 
              & 1971.48     & 0.222       &                       \\
 RCHB (PC-PK1) \cite{Xia2018_ADNDT121-122-1} \tnote{1)}          
              & 1952.65     &             & 6.132                 \\ 
 RCHB (PC-PK1) + RBF \cite{Shi2019_ChinPhysC_in-press} \tnote{2)}          
              & 1969.20     &             & 6.132                 \\ 
\bottomrule
\end{tabular}
\begin{tablenotes}
 \item[1)] In Ref.~\cite{Xia2018_ADNDT121-122-1}, the spherical symmetry is 
           assumed for all nuclei.
 \item[2)] In Ref.~ \cite{Shi2019_ChinPhysC_in-press}, 
           the RBF approach was applied to nuclear masses but not to charge radii 
           predicted by the RCHB model in Ref.~\cite{Xia2018_ADNDT121-122-1}.
\end{tablenotes}
\end{threeparttable}
\end{table}

The ground state binding energy $E_\mathrm{B}$ of $^{270}$Hs from our MDC-RMF model 
calculation is 1967.40 MeV. 
There is no experimental value for the binding energy of $^{270}$Hs and in 
AME2016 $E_\mathrm{B}$ = 1969.65 MeV was derived from the trends in 
the mass surface (TMS) \cite{Audi2017_ChinPhysC41-030001,
Huang2017_ChinPhysC41-030002,Wang2017_ChinPhysC41-030003}.
Our calculation result is very close to this value.
In fact, almost all mass models, including the RMF models with various effective
interactions \cite{Ren2002_PRC65-051304R,Ren2002_PRC66-064306,
Geng2005_PTP113-785,Geng2006_PhD-Thesis,
Zhang2012_PRC85-014325}, the Skyrme Hartree--Fock--Bogoliubov
mass model (HFB-24) \cite{Goriely2013_PRC88-024308},
and the three typical macroscopic-microscopic models (MMM)---
the one by Patyk et al. \cite{Patyk1991_NPA533-132},
the Weizs\"acker--Skyrme model \cite{Wang2010_PRC81-044322,Wang2010_PRC82-044304,
Liu2011_PRC84-014333} (WS4 \cite{Wang2014_PLB734-215}) and 
the finite range droplet model [FRDM (2012)] \cite{Moeller2016_ADNDT109-110-1},
predicted similar $E_\mathrm{B}$ values as is seen in Table~\ref{table:g.s.&comp}. 
The only exception is the relativistic continuum Hartree--Bogoliubov (RCHB) 
model \cite{Meng1996_PRL77-3963,Meng1998_PRL80-460,Meng1998_NPA635-3}
which has been used to study how the continuum effects can extend the nuclear 
chart \cite{Qu2013_SciChinaPMA56-2031,Xia2018_ADNDT121-122-1}.
The binding energy calculated for $^{270}$Hs from this model is 1952.65 MeV
\cite{Xia2018_ADNDT121-122-1} which is much smaller than those from other model
predictions and the empirical value from AME2016.
The reason is that in the RCHB model, the spherical symmetry is assumed for
all nuclei and this is certainly not true for $^{270}$Hs.
However, it is interesting to see that the radial basis function (RBF) approach
\cite{Buhmann2006_RBF,Wang2011_PRC84-051303R,Zheng2014_PRC90-014303,Niu2016_PRC94-054315}
can eliminate local systematic deviations between the experimental mass values
and those calculated from the RCHB model, thus improving very much the 
predictive power of the RCHB model for nuclear masses 
\cite{Shi2019_ChinPhysC_in-press}.
For $^{270}$Hs, the binding energy from the RCHB+RBF approach is 1969.20 MeV
which is fairly close to the AME2016 value as well as other model predictions.
Note that a deformed relativistic Hartree--Bogoliubov model in continuum 
(the DRHBc model) has been developed and used to study the deformation and continuum 
effects in exotic nuclei \cite{Zhou2010_PRC82-011301R,Li2012_PRC85-024312,
Li2012_CPL29-042101,Sun2018_PLB785-530}. 
An international collaboration 
is working on producing a new chart of nuclides and a new mass table with 
the DRHBc model \cite{DRHBc-mass-table-collaboration}.

There are no experimental information on the quadrupole moment and deformation 
parameter $\beta_2$ for $^{270}$Hs.
From all model calculations with the deformation considered, $^{270}$Hs 
is well deformed in the ground state.
From our MDC-RMF calculation, the quadrupole deformations of the ground state
of $^{270}$Hs are $\beta_{20} = 0.261$ and $\beta_{22} = 0$, meaning that 
$\beta_2 = 0.261$. 
This $\beta_2$ value is quite similar to the results from HFB-24 and RMF model 
calculations with NLZ2 and NL3, but is larger than the RMF results with TMA and MMMs.
While in the MDC-RMF model the reflection asymmetric shapes are allowed,
the calculated octupole deformation $\beta_{30} = 0$ 
for the ground state of $^{270}$Hs. 
The calculated hexadecupole deformation $\beta_{40} = -0.056$ which is 
comparable to
$\beta_{40} = -0.04$ from HFB-24 \cite{Goriely2013_PRC88-024308},
$\beta_{40} = -0.052$ from WS4 \cite{Wang2014_PLB734-215} and
$\beta_{40} = -0.079$ from the FRDM (2012) \cite{Moeller2016_ADNDT109-110-1}.
The rms charge radius $R_\mathrm{c}$ has not been measured for $^{270}$Hs. 
All calculations including ours give predictions of $R_\mathrm{c}$ values between 
6.13 fm and 6.26 fm.
Note that in Ref.~\cite{Shi2019_ChinPhysC_in-press}, the RBF approach was not 
applied to the charge radii obtained from the RCHB calculations. 

\begin{figure}[H]
\centering{\includegraphics[width=0.98\columnwidth]{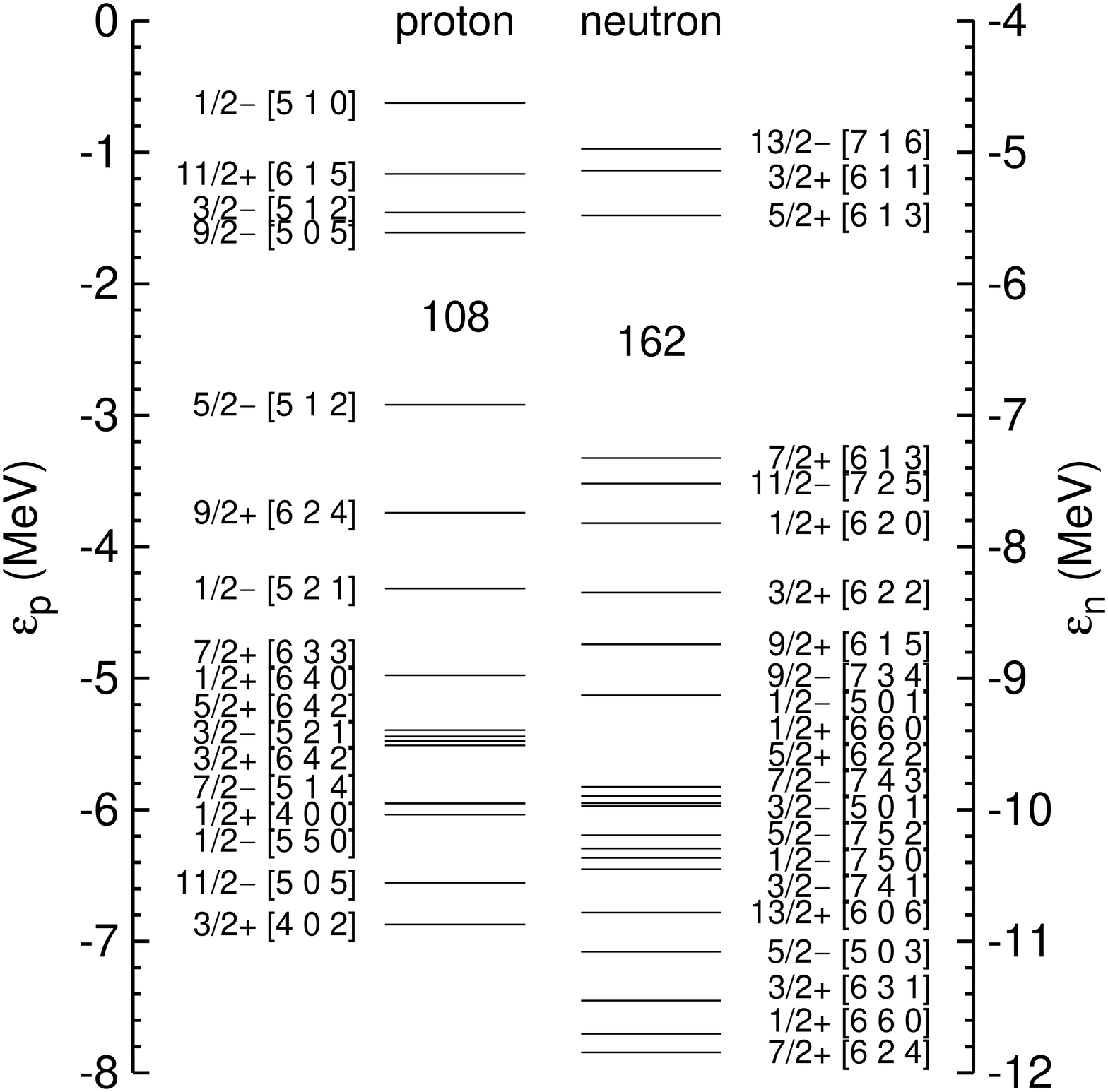}}
\caption{(Color online) 
Single proton and neutron levels of the ground state of $^{270}$Hs.
The Nilsson quantum numbers are given for each level.
}
\label{fig:ground state sp-level}
\end{figure}

Single proton and neutron energy levels of $^{270}$Hs in the ground state 
are shown in \autoref{fig:ground state sp-level}.
Since $^{270}$Hs is an axially deformed nucleus with the reflection symmetry, 
we can label each single particle level with the Nilsson quantum numbers.
In \autoref{fig:ground state sp-level}, one can see clearly the feature of
the double magicity. 
The shell gap is 1.3 MeV for protons at $Z=108$ and around 2 MeV 
for neutrons at $N=162$.
Such gaps are very prominent considering the fact that in the superheavy region 
the single particle level density is quite large and grows faster than expected 
from the $A^{1/3}$ scaling even for spherical SHN \cite{Agbemava2015_PRC92-054310}.
In Ref.~\cite{Patyk1991_NPA533-132}, the $Z=108$ shell gap is similar as
that in the present work, while the energy gap at $N=162$ is
a bit smaller than ours.

\subsection{One-dimensional potential energy curves of $^{270}$Hs}\vspace*{-1mm}

In this subsection, we present the results of the MDC-RMF calculations with a
constraint made on the axial quadrupole deformation $\beta_{20}$ for $^{270}$Hs. 
We examine one-dimensional (1D) PECs and discuss the influences of the axial 
octupole deformation $\beta_{30}$ and non-axial quadrupole deformation 
$\beta_{22}$ along these PECs.
The (dis)continuities of these PECs are investigated in detail.

In the MDC-RMF model, both the triaxial (TA) and reflection asymmetric (RA) 
deformations are allowed. 
We may switch off TA or RA deformations and keep the nucleus
in question to be axially symmetric (AS) or reflection symmetric (RS).
Thus four typical combinations of symmetries can be imposed in the MDC-RMF 
calculations: AS-RS, AS-RA, TA-RS and TA-RA.
In the 1D constraint calculations in the present work, we take
$\beta_{20}$ values running from 0 to 1.5 with the step size 
$\delta\beta_{20}$ = 0.02 when $\beta_{20} < 1$. 
When $\beta_{20} > 1$, we take $\delta\beta_{20} = 0.02$ in AS-RS calculations
and $\delta\beta_{20} = 0.05$ in the AS-RA, TA-RS and TA-RA calculations.

\begin{figure}[H]
\centering{\includegraphics[width=0.98\columnwidth]{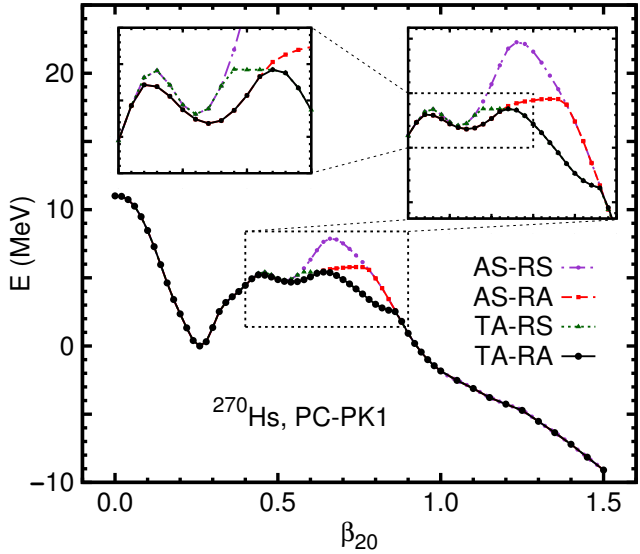}}
\caption{(Color online) 
Potential energy curves of $^{270}$Hs obtained from MDC-RMF calculations 
with four combinations of symmetries imposed:
axially symmetric and reflection symmetric (AS-RS) deformations, 
axially symmetric and reflection asymmetric (AS-RA) deformations, 
triaxial and reflection symmetric (TA-RS) deformations, and 
triaxial and reflection asymmetric (TA-RA) deformations.
The inset on the top right shows these four PECs in the region 
$0.4 \leq \beta_{20} \leq 0.9$ beyond which the four curves are identical.
The inset on the top left shows these PECs in the region 
$0.4 \leq \beta_{20} \leq 0.7$.
Note that the TA-RS PEC overlaps with the TA-RA curve when $\beta_{20} \geq 0.64$
and the AS-RA and TA-RA PECs overlap with each other when $\beta_{20} \leq 0.62$.
\label{fig:1d_PEC}}
\end{figure}

Four PECs of $^{270}$Hs calculated from the MDC-RMF model are 
shown in \autoref{fig:1d_PEC}. 
These four PECs are identical in the following $\beta_{20}$ intervals:
$0 \le \beta_{20} \le 0.42$ and $0.88 \le \beta_{20} \le 1.50$.
In particular, the global minimum corresponding to the ground state of 
$^{270}$Hs coincides in all these four calculations regardless of the 
imposed symmetries.
For the AS-RS PEC in \autoref{fig:1d_PEC}, the first fission barrier occurs at 
around $\beta_{20}=0.46$ and the height of this barrier is 5.4 MeV.
The second minimum in the AS-RS PEC occurs at $\beta_{20}=0.52$ and the energy 
of this minimum is 4.8 MeV above the ground state. 
The depth of the pocket around the second minimum is about 0.6 MeV.
The second fission barrier is around $\beta_{20}=0.66$ with a height of 7.9 MeV.
If the reflection asymmetry is considered, one gets the AS-RA PEC in which
the height of the first barrier is 5.2 MeV, which is by 0.2 MeV lower than that in
the AS-RS curve. 
The second minimum is shifted to $\beta_{20}=0.54$ and the energy is by 0.1 MeV
smaller than that from the reflection symmetric calculation. 
Though it does not affect the first barrier and the second minimum very much, 
the reflection asymmetric shape has a considerably large effect on the second
fission barrier which is not only shifted to $\beta_{20}=0.74$ but also 
lowered by 2.1 MeV (cf. violet and red curves in \autoref{fig:1d_PEC}). 
The TA-RS PEC is still of double-hump structure and the depth of the pocket
around the second minimum is about 0.6 MeV.
It can be seen that the triaxial distortion does not influence much the first 
barrier and the second minimum, but lowers the second fission barrier 
by about 2.4 MeV.
This lowering effect is very pronounced considering that the 
height of the second barrier in the AS-RS PEC is only 7.9 MeV.
The fact that the triaxiality affects much only the second fission barrier in $^{270}$Hs
is different from the findings in actinides in which the triaxial deformations
lowers both the first and the second fission barriers considerably
\cite{Lu2012_PRC85-011301R,Lu2014_PRC89-014323}.
When both triaxial and reflection asymmetric shapes are allowed in the 
MDC-RMF calculations, we obtain the TA-RA PEC which is shown in 
\autoref{fig:1d_PEC} as black dots connected with a solid line. 
Compared with the TA-RS PEC, the TA-RA curve is slightly lower in the region 
$0.44 \le \beta_{20} \le 0.62$ as can be seen more clearly in the insets 
of \autoref{fig:1d_PEC}.
We note that the RS PECs have been obtained for $^{270}$Hs from 
Woods--Saxon--Strutinsky calculations in reflection symmetric deformation space 
$(\beta_2,\gamma,\beta_4)$ \cite{Chai2018_ChinPhysC42-54101}.

\begin{figure}[H]
\centering{\includegraphics[width=0.88\columnwidth]{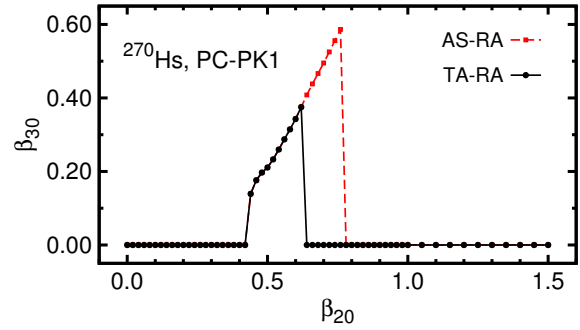}}
\caption{(Color online) 
Axial octupole deformation parameter $\beta_{30}$ as a function of 
$\beta_{20}$ along the axially symmetric and reflection asymmetric (AS-RA) and 
triaxial and reflection asymmetric (TA-RA) potential energy curves shown 
in \autoref{fig:1d_PEC}. 
Note that the two curves overlap with each other when 
$\beta_{20} \leq 0.62$ and $\beta_{20} \geq 0.78$.
\label{fig:beta_30-curve}}
\end{figure}

\begin{figure}[H]
\centering{\includegraphics[width=0.88\columnwidth]{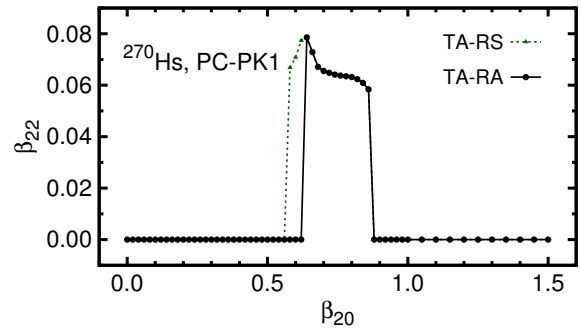}}
\caption{(Color online) 
Triaxial deformation parameter $\beta_{22}$ as a function of 
$\beta_{20}$ along the triaxial and reflection symmetric (TA-RS) and 
triaxial and reflection asymmetric (TA-RA) potential energy curves shown 
in \autoref{fig:1d_PEC}. 
Note that the two curves overlap with each other when 
$\beta_{20} \leq 0.56$ and $\beta_{20} \geq 0.64$.
}\label{fig:beta_22-curve}
\end{figure}

In \autoref{fig:1d_PEC}, the energy is continuous as a function of $\beta_{20}$ 
along the four PECs.
However, the $\beta_{30} \sim \beta_{20}$ and $\beta_{22} \sim \beta_{20}$ 
curves are discontinuous along the AS-RA, TA-RS and TA-RA PECs as shown in 
\autoref{fig:beta_30-curve} and \autoref{fig:beta_22-curve}.
In \autoref{fig:beta_30-curve}, we can see that $\beta_{30}$ keeps to be 0 
when $0 \leq \beta_{20} \leq 0.42$, jumps to 0.14 at 
$\beta_{20} = 0.44$ and increases monotonically until $\beta_{20}=0.76$ 
in the AS-RA curve. 
After a sudden drop at $\beta_{20} = 0.78$, $\beta_{30}$ becomes 0 again. 
In the TA-RS PEC shown in \autoref{fig:beta_22-curve}, $\beta_{22}$ 
is 0 until $\beta_{20} = 0.56$. 
Then $\beta_{22}$ jumps to 0.067 at $\beta_{20} = 0.58$ and 
changes continuously in the region $0.58 \leq \beta_{20} \leq 0.86$.
At $\beta_{20} = 0.88$, $\beta_{22}$ drops again to 0. 
As for the TA-RA PEC, since both triaxial and reflection asymmetric shapes
are allowed, both $\beta_{22}$ and $\beta_{30}$ can be non-zero.
It is seen that in \autoref{fig:beta_30-curve} there are sudden changes in
$\beta_{30}$ at $\beta_{20} = 0.44$ and $\beta_{20} = 0.64$ while in 
\autoref{fig:beta_22-curve} there are sudden changes in $\beta_{22}$ 
at $\beta_{20} = 0.64$ and $\beta_{20} = 0.88$. 
Moreover, in the TA-RA PEC obtained by constraining only $\beta_{20}$,
either $\beta_{22}$ or $\beta_{30}$ is 0 which means that the triaxial 
and octupole deformations do not co-exist.
Note that in actinide nuclei the triaxial and octupole deformations
do co-exist around the second fission barriers 
\cite{Lu2012_PRC85-011301R,Lu2014_PRC89-014323}.

These sudden changes of $\beta_{30}$ or $\beta_{22}$ in the AS-RA, TA-RS and TA-RA 
PECs come from the complexity of multidimensional PESs 
\cite{Moeller2009_PRC79-064304,Lu2014_PRC89-014323,Dubray2012_CPC183-2035}. 
In the self-consistent calculations constraining only $\beta_{20}$, shape 
degrees of freedom other than $\beta_{20}$ are minimized automatically in 
the process of finding (local) minima in a PES. 
It often happens that there are two or more valleys in a higher-dimensional PES.
When such a PES is projected onto a lower-dimensional one, 
these valleys may overlap or be connected abruptly, 
resulting in ``continuous'' PESs 
(e.g., the AS-RA, TA-RS and TA-RA PECs in \autoref{fig:1d_PEC}) 
though the deformation parameters which are projected out 
may change suddenly at intersecting or connecting point(s)
(e.g., $\beta_{30}$ in \autoref{fig:beta_30-curve} 
or $\beta_{22}$ in \autoref{fig:beta_22-curve}). 
In fact, the AS-RA or TA-RS PECs shown in \autoref{fig:1d_PEC} consist of 
three isolated curves in the two-dimensional PES; e.g., for the AS-RA PEC,
the three parts correspond to the following three $\beta_{20}$ intervals
as seen in \autoref{fig:beta_30-curve}:
$(0 \leq \beta_{20} \leq 0.42)$, $(0.44 \leq \beta_{20} \leq 0.76)$
and $(0.78 \leq \beta_{20} \leq 1.50)$. 
As seen in \autoref{fig:beta_30-curve} and \autoref{fig:beta_22-curve},
the TA-RA PEC in \autoref{fig:1d_PEC} consists of even four segments in the
three-dimensional (3D) coordinate system with $\beta_{20}$, $\beta_{22}$,
and $\beta_{30}$ as coordinates:
$(0 \leq \beta_{20} \leq 0.42)$, $(0.44 \leq \beta_{20} \leq 0.62)$,
$(0.64 \leq \beta_{20} \leq 0.86)$ and $(0.88 \leq \beta_{20} \leq 1.50)$. 
The first and forth segments are along the $\beta_{20}$--axis 
($\beta_{30}=0$, $\beta_{22}=0$), the second 
is in the $\beta_{20}$--$\beta_{30}$ plane ($\beta_{30} \ne 0$, $\beta_{22}=0$) and 
the third is in the $\beta_{20}$--$\beta_{22}$ plane ($\beta_{30}=0$, $\beta_{22}\ne0$).
It is clear that one needs to examine higher-dimensional PESs in order to 
eliminate such discontinuities.

\subsection{Two-dimensional potential energy surfaces of $^{270}$Hs 
\label{subsec:2d_PES}}

In this subsection, we present and discuss the two-dimensional (2D) PESs of 
$^{270}$Hs which are obtained by simultaneously constraining 
the axial quadrupole deformation $\beta_{20}$ and 
the axial octupole deformation $\beta_{30}$.
In the 2D constraint calculations, we take $\beta_{20}$ values running 
from 0.00 to 2.00 and $\beta_{30}$ from 0.00 to 3.00, both with a step size of 0.02. 
One certainly does not need to calculate all points on this deformation lattice 
thus defined. 
The points in the top left corner of the PES with energy larger than 16 MeV 
(relative to the ground state) and those in the bottom right corner with energy 
lower than the ground state are not calculated.

\begin{figure}[H]
\centering{\includegraphics[width=0.98\columnwidth]{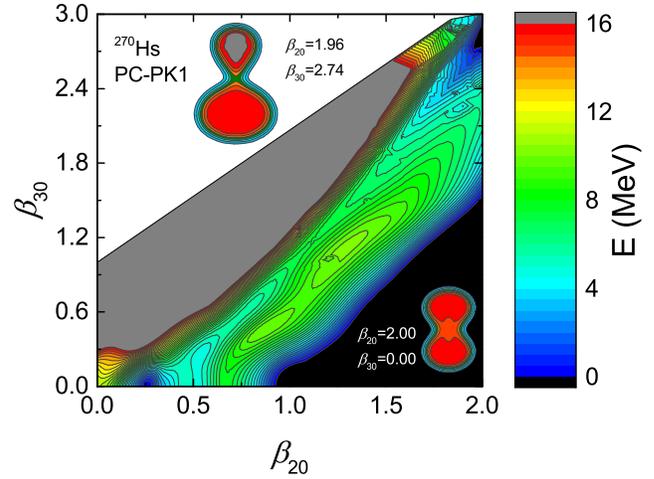}}
\caption{(Color online) 
Potential energy surface of $^{270}$Hs obtained from 2D constraint RMF
calculations with axial and reflection asymmetric (AS-RA) shapes allowed.
The energy is normalized with respect to the binding energy of the ground state.
The contour interval is 0.5 MeV.
The density profiles of $^{270}$Hs at $(\beta_{20} = 1.96,\ \beta_{30} = 2.74)$
and $(\beta_{20} = 2.00,\ \beta_{30} = 0)$ are shown.
\label{fig:2D-PES_ASRA}}
\end{figure}

The 2D PES for $^{270}$Hs with the axial symmetry is shown in \autoref{fig:2D-PES_ASRA}. 
This PES is obtained from the MDC-RMF calculations with $\beta_{20}$ and 
$\beta_{30}$ constrained and the axial symmetry imposed.
In \autoref{fig:2D-PES_ASRA}, one can find two valleys corresponding to
two possible fission paths. 
One of them goes along the axis of abscissas, i.e., $^{270}$Hs keeps $\beta_{30} = 0$
when it is elongated along this fission path. 
The energy of $^{270}$Hs is smaller than the ground state energy when 
$\beta_{20} \approx 0.9$. 
$^{270}$Hs becomes very soft against $\beta_{30}$ distortion for 
$\beta_{20} > 0.4$ where the other possible fission path appears along the 
reflection asymmetric valley.
The reflection asymmetric fission path extends much farther than the reflection
symmetric one in the directions of both $\beta_{20}$ and $\beta_{30}$ and
the energy goes below the ground state at $\beta_{20} \approx 1.96$ and
$\beta_{30} \approx 2.7$. 
In \autoref{fig:2D-PES_ASRA}, the density profiles of $^{270}$Hs corresponding 
to the two fission configurations are also shown. 
The two fragments in the symmetric fission can certainly be identified as $^{135}$Xe.
However, the fission along the RA path is extremely asymmetric: 
The heavy fragment corresponds to $^{208}$Pb and the light one to $^{62}$Fe.
Such an asymmetric fission may be considered as cluster radioactivity caused by 
the strong shell effects in $^{208}$Pb \cite{Matheson2018_arXiv1812.06490}.

\begin{figure}[H]
\centering{\includegraphics[width=0.88\columnwidth]{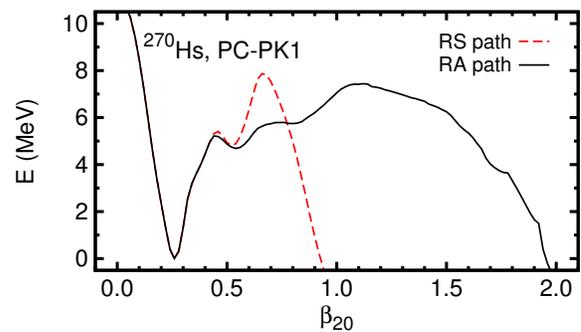}}
\caption{(Color online) 
Potential energy curves of $^{270}$Hs corresponding to the reflection 
symmetric (RS) and reflection asymmetric (RA) fission paths in 
\autoref{fig:2D-PES_ASRA}. 
The energy is normalized with respect to the binding energy of the ground state.
Note that the two curves overlap with each other when $\beta_{20} < 0.44$.
\label{fig:two-paths_AS}}
\end{figure}

Along both paths, there are two or more minima and barriers which can be seen
more clearly in \autoref{fig:two-paths_AS}.
The RS path in \autoref{fig:two-paths_AS} corresponds to the AS-RS PEC in
\autoref{fig:1d_PEC}.
At $\beta_{20} = 0.78$ the RS and RA PECs cross in \autoref{fig:two-paths_AS}.
When $\beta_{20} \leq 0.76$, the RA PEC in \autoref{fig:two-paths_AS} is the 
same as the AS-RA PEC in \autoref{fig:1d_PEC}, 
while it is much higher than the latter when $\beta_{20} \geq 0.78$. 
The energy minimization in 1D constraint RMF calculations makes the AS-RA curve 
in \autoref{fig:1d_PEC} follow the valley with the lowest energy which 
corresponds to the RS path with $\beta_{20} \geq 0.78$ in 
\autoref{fig:two-paths_AS}.
Only if 2D constraints on both $\beta_{20}$ and $\beta_{30}$ are made,
one can stick to the reflection asymmetric path even though energies of
some points in this path are not the lowest.
Along the RA path, one can find a pocket around $\beta_{20} = 0.54$ 
($\beta_{30} = 0.26$ as read in \autoref{fig:2D-PES_ASRA}) 
and a shoulder around $\beta_{20} = 0.80$ ($\beta_{30} = 0.64$).
The highest barrier with a height 7.4 MeV is around 
$\beta_{20} = 1.12$ ($\beta_{30} = 1.18$),
which is lower by about 0.5 MeV than the barrier around $\beta_{20} = 0.66$ 
in the RS path.

\begin{figure}[H]
\centering{\includegraphics[width=0.98\columnwidth]{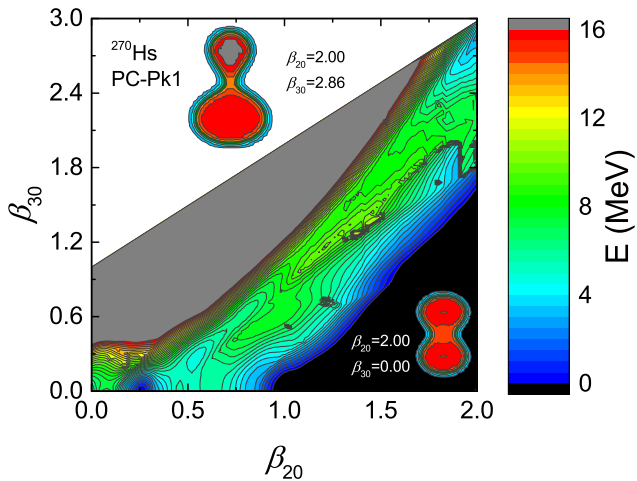}}
\caption{(Color online) 
Potential energy surface of $^{270}$Hs obtained from 2D constraint RMF
calculations with triaxial and reflection asymmetric (TA-RA) deformations allowed.
The energy is normalized with respect to the binding energy of the ground state.
The contour interval is 0.5 MeV.
The density profiles of $^{270}$Hs at $(\beta_{20} = 2.00,\ \beta_{30} = 2.86)$
and $(\beta_{20} = 2.00,\ \beta_{30} = 0)$ are shown.
\label{fig:2D-PES_TARA}}
\end{figure}

\begin{figure}[H]
\centering{\includegraphics[width=0.88\columnwidth]{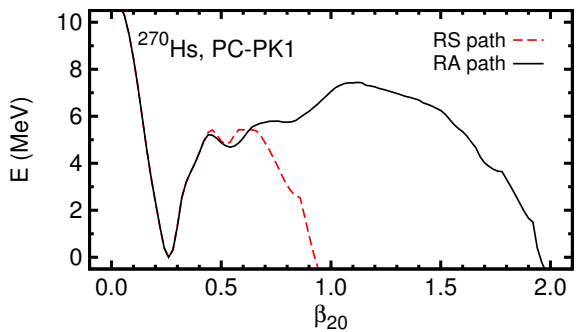}}
\caption{(Color online) 
Potential energy curves of $^{270}$Hs corresponding to the reflection
symmetric (RS) and reflection asymmetric (RA) fission paths in 
\autoref{fig:2D-PES_TARA} but recalculated with the ADHO basis truncated with $N_f=20$. 
The energy is normalized with respect to the binding energy of the ground state.
Note that the two curves overlap with each other when $\beta_{20} < 0.44$.
\label{fig:two-paths_TA}}
\end{figure}

2D-constraint RMF calculations are carried out for $^{270}$Hs with both
triaxial and reflection asymmetric shapes allowed. 
The 2D PES thus obtained is shown in \autoref{fig:2D-PES_TARA}. 
The TA PES shares some common features as the AS PES in \autoref{fig:2D-PES_ASRA}.
There are also two competing fission paths, the RS path along the $\beta_{20}$-axis 
with $\beta_{30} = 0$ and the RA path going to the top right corner of this plot.
In this PES, there are some discontinuities, e.g., around 
$(\beta_{20} \approx 1.2,\  \beta_{30} \approx 0.7)$,
$(\beta_{20} \approx 1.5,\  \beta_{30} \approx 1.4)$
and $(\beta_{20} \approx 1.9,\ \beta_{30} \approx 1.8)$.
These discontinuities may also be due to the complexity of multidimensional PES, 
similar to those in 1D PECs as shown in \autoref{fig:beta_30-curve}
and \autoref{fig:beta_22-curve}.
To resolve these discontinuities, one has to investigate 3D PES by making 3D 
constraint calculations on $\beta_{20}$, $\beta_{30}$ and $\beta_{22}$.

By comparing \autoref{fig:2D-PES_ASRA} and \autoref{fig:2D-PES_TARA} and 
examining the $\beta_{22}$ values of the points in the PES shown in 
\autoref{fig:2D-PES_TARA}, it is found that the triaxial deformation is zero 
along the reflection asymmetric fission path. 
However, 
the triaxial effects are significant along the reflection symmetric fission path.
In \autoref{fig:two-paths_TA}, the PECs of $^{270}$Hs corresponding to the RS 
and RA fission paths in \autoref{fig:2D-PES_TARA} are presented.
To discuss quantitatively the influence of the triaxiality, these two
curves are obtained from MDC-RMF calculations with the ADHO basis space truncated 
up to $N_f=20$. 
In fact, the RA path in \autoref{fig:two-paths_TA} (the black curve) is the same 
as that in \autoref{fig:two-paths_AS} because $^{270}$Hs does not favor a triaxial 
deformation along this path.
Moreover, the RS PEC in \autoref{fig:two-paths_TA} is the same 
as the TA-RS PEC in \autoref{fig:1d_PEC}.
In contrast with the RA path, the RS fission path is lowered considerably by
the triaxial distortion. 
Compared to the RS path in \autoref{fig:two-paths_AS},  
the second barrier of the RS fission path in \autoref{fig:two-paths_TA} is 
lowered by 2.4 MeV.
As a result, the fission barrier along the RS path is much lower than 
that in the RA path.
Comparing \autoref{fig:1d_PEC}, \autoref{fig:two-paths_AS} and 
\autoref{fig:two-paths_TA}, we finally come to the conclusion that
the TA-RS PEC corresponds to the lowest static fission path though it
is higher in energy than the TA-RA curve in \autoref{fig:1d_PEC}.
Such a conclusion can only be reached when both 1D PECs and 2D PESs are 
examined carefully and when both triaxial and reflection asymmetric deformations
are considered.

\section{Summary and perspectives} \vspace*{-1mm}
\label{sec:summary}

We have investigated in detail the doubly magic deformed SHN
$^{270}$Hs with the MDC-RMF model. The successful effective interaction PC-PK1 
is adopted for the covariant density functional and a separable pairing force 
of finite range is used in the pp channel. 
The binding energy, rms charge radius and deformation parameters obtained for
the ground state of this nucleus are compared with predictions from other
models and the empirical value in AME2016 and good agreements are found. 
In single particle level schemes of protons and neutrons, there are 
relatively large shell gaps at $Z=108$ and $N=162$ which cause the 
deformed double magicity in $^{270}$Hs. 

If only 1D PECs are examined, one is easily led to a wrong static fission path 
which is, though the lowest in energy, very likely to be discontinuous in 2D 
or higher-dimensional PESs.
Therefore we investigate in detail 1D PECs and 2D PESs with various spatial 
symmetries imposed.
We stress that it is for the first time the PES is presented for $^{270}$Hs 
with both triaxiality and reflection asymmetry considered.
When the axial symmetry is assumed, the reflection asymmetric fission barrier 
is lower than the reflection symmetric one. 
When the triaxial deformation is allowed, the reflection symmetric fission path
becomes favorable because the reflection symmetric barrier is lowered very much
by the triaxial distortion.
Although it is higher than the reflection symmetric one, the reflection 
asymmetric fission barrier may still be penetrated or overcome by $^{270}$Hs in
spontaneous or induced fissions. If this happens, $^{270}$Hs would end with 
fragments with very large mass asymmetry, $^{208}$Pb and $^{62}$Fe, which 
may be considered as cluster emission resulted from the strong shell effects.

The above conclusions will be useful for learning detailed information about 
PESs and fission barrier properties of SHN. 
Furthermore, as the first thorough study on SHN with the MDC-RMF model, 
this work serves as the starting point of systematic investigations of SHN 
with MDC-CDFTs. 
In the future, more efforts will be devoted to the calculations of TA-RA PESs 
with larger $N_f$ and 3D PESs, systematic studies of static fission paths and 
fission barriers of SHN, 
investigations of fission dynamics based on the PESs from MDC-CDFTs and
cluster radioactivity, etc.

\Acknowledgements{
We thank Zhong-Ming Niu for allowing us to use the results for $^{270}$Hs in 
Ref.~\cite{Shi2019_ChinPhysC_in-press} prior to publication and discussions
concerning nuclear binding energy and mass excess.
This work was supported by 
the National Key R\&D Program of China (Grant No. 2018YFA0404402), 
the National Natural Science Foundation of China (Grants 
No. 11525524, No. 11621131001, No. 11647601, No. 11747601, and No. 11711540016),
the Key Research Program of Frontier Sciences of Chinese Academy of Sciences (No. QYZDB-SSWSYS013),
the Key Research Program of Chinese Academy of Sciences (No. XDPB09-02),
the Inter-Governmental S\&T Coorperation Project between China and Croatia,
and 
the IAEA Coordinated Research Project ``F41033''. 
The results described in this paper are obtained on 
the High-performance Computing Cluster of KLTP/ITP-CAS and
the ScGrid of the Supercomputing Center, Computer Network Information Center of Chinese Academy of Sciences.
}

\InterestConflict{The authors declare that they have no conflict of interest.}





\end{multicols}
\end{document}